\documentclass[aps,prl,amsmath,amssymb,floatfix,twocolumn,amsmath,superscriptaddress,twocolumn,nofootinbib,tighten,letterpaper]{revtex4}
\usepackage{multirow}
\usepackage{bbold}
\usepackage{subfigure}
\usepackage{color}
\usepackage{xcolor}
\usepackage{mathrsfs}
\usepackage{hyperref}
\usepackage{bm}
\usepackage{stackrel}

\usepackage{amsfonts, relsize, color}
\usepackage{graphics}
\usepackage{graphicx}
\usepackage{amssymb}
\usepackage{amsbsy}
\usepackage{amsmath}
\usepackage{array}
\usepackage{fontenc}
\usepackage{textcomp}
\usepackage{rotating}
\usepackage{comment}
\usepackage{braket}

\usepackage{notes2bib}

\usepackage{calc}
\usepackage{tikz}
\usetikzlibrary{math}
\newlength{\nsubht}
\newsavebox{\nsubbox}

\newcommand{\sx}{\sin(k_x a)}
\newcommand{\sy}{\sin(k_y a)}
\newcommand{\sz}{\sin(k_z a)}
\newcommand{\cx}{\cos(k_x a)}
\newcommand{\cy}{\cos(k_y a)}
\newcommand{\cz}{\cos(k_z a)}

\newcommand{\Sx}{\sin(k_x)}
\newcommand{\Sy}{\sin(k_y)}

\newcommand{\Cx}{\cos(k_x)}
\newcommand{\Cy}{\cos(k_y)}

\newcommand{\TKIpicture}[3]{%
	\sbox\nsubbox{
		\resizebox{\columnwidth}{!}
		{%
			\includegraphics[height=3cm]{#1}%
			\includegraphics[height=3cm]{#2}%
			\includegraphics[height=3cm]{#3}%
		}%
	}%
	\setlength{\nsubht}{\ht\nsubbox}%
	\begin{tikzpicture}[scale=1]
	\node[inner sep=0pt,anchor=west] (NOOP) at (0,0) %
	{\includegraphics[height=\nsubht]{#1}};
	\node[inner sep=0pt,anchor=west] (OP1) at (NOOP.east) %
	{\includegraphics[height=\nsubht]{#2}};
	\node[inner sep=0pt,anchor=west] (cbar) at (OP1.east) %
	{\includegraphics[height=\nsubht]{#3}};
	
	\node (NOOPcap) at ([xshift=-0.7\nsubht,yshift=-0.1\nsubht]NOOP.north east) %
	{(a)};
	\node (OP1cap) at ([xshift=-0.7\nsubht,yshift=-0.1\nsubht]OP1.north east) %
	{(b)};
	\end{tikzpicture}%
}

\newcommand{\diracpicture}[4]{%
	\sbox\nsubbox{
		\resizebox{0.9\columnwidth}{!}
		{%
			\includegraphics[height=3cm]{#1}%
			\includegraphics[height=3cm]{#2}%
		}%
	}%
	\setlength{\nsubht}{\ht\nsubbox}%
	\begin{tikzpicture}[scale=1]
		\node[inner sep=0pt,anchor=west] (NOOP) at (0,0) %
		{\includegraphics[height=\nsubht]{#1}};
		\node[inner sep=0pt,anchor=west] (OP1) at (NOOP.east) %
		{\includegraphics[height=\nsubht]{#2}};
		\node[inner sep=0pt,anchor=west] (OP2) at (0,-0.9\nsubht) %
		{\includegraphics[height=\nsubht]{#3}};
		\node[inner sep=0pt,anchor=west] (cbar) at ([yshift=-0.5\nsubht+\nsubht-0.9\nsubht]OP1.south) %
		{\includegraphics[height=\nsubht]{#4}};
		
		\node (NOOPcap) at ([xshift=-0.7\nsubht,yshift=-0.1\nsubht]NOOP.north east) %
		{(a)};
		\node (OP1cap) at ([xshift=-0.7\nsubht,yshift=-0.1\nsubht]OP1.north east) %
		{(b)};
		\node (OP2cap) at ([xshift=-0.7\nsubht,yshift=-0.125\nsubht]OP2.north east) %
		{(c)};
	\end{tikzpicture}%
}
\newcommand{\LNSMpicture}[8]{%
		\sbox\nsubbox{
			\resizebox{0.9\textwidth}{!}
			{%
				\includegraphics[scale=1]{#1}%
				\includegraphics[scale=1]{#2}%
				\includegraphics[scale=1]{#3}%
				\includegraphics[scale=1]{#4}%
			}%
		}%
		\setlength{\nsubht}{\ht\nsubbox}%
		\centering%
		\begin{tikzpicture}
			\tikzmath
			{
				\imSize=183;
				\padDown=48;
				\cbarLeft=5;
				\padUp=10;
				\yOff1=(\padUp+0.15*(\imSize-\padDown-\padUp))/\imSize*\nsubht;
				\xOff1=(\padUp+0.85*(\imSize-\padDown-\padUp))/\imSize*\nsubht;
				\yOff2=(\padUp+0.5*(\imSize-\padDown-\padUp))/\imSize*\nsubht;
				\xOff2=(\padUp+0.5*(\imSize-\padDown-\padUp))/\imSize*\nsubht;
				\cbarXoff=(\padDown-\cbarLeft)/\imSize*\nsubht;
			}
		\node[inner sep=0pt,anchor=west] (NOOPk) at (0,0) %
		{\includegraphics[height=\nsubht]{#1}};
		\node[inner sep=0pt,anchor=west] (OP2k) at (NOOPk.east) %
		{\includegraphics[height=\nsubht]{#2}};
		\node[inner sep=0pt,anchor=west] (OP23k) at (OP2k.east) %
		{\includegraphics[height=\nsubht]{#3}};
		\node[inner sep=0pt,anchor=west] (OP123k) at (OP23k.east) %
		{\includegraphics[height=\nsubht]{#4}};
		
		\node[inner sep=0pt,anchor=north] (NOOP) at (NOOPk.south) %
		{\includegraphics[height=\nsubht]{#5}};
		\node[inner sep=0pt,anchor=west] (OP2) at (NOOP.east) %
		{\includegraphics[height=\nsubht]{#6}};
		\node[inner sep=0pt,anchor=west] (OP23) at (OP2.east) %
		{\includegraphics[height=\nsubht]{#7}};
		\node[inner sep=0pt,anchor=east] (cbar) at ([yshift=-\nsubht]OP123k.east) %
		{\includegraphics[height=\nsubht]{#8}};		
		\node (NOOPkcap) at
		([xshift=-\xOff1,yshift=-\yOff1]NOOPk.north east) %
		{\color{white}(a1)};
		\node (OP2kcap) at
		([xshift=-\xOff1,yshift=-\yOff1]OP2k.north east) %
		{\color{white}(a2)};
		\node (OP23kcap) at
		([xshift=-\xOff1,yshift=-\yOff1]OP23k.north east) %
		{\color{white}(a3)};    	
		\node (OP123kcap) at
		([xshift=-\xOff1,yshift=-\yOff1]OP123k.north east) %
		{\color{white}(a4)};    	
		\node (NOOPcap) at
		([xshift=-\xOff2,yshift=-\yOff2]NOOP.north east) %
		{\color{red}(b1)};
		\node (OP2cap) at
		([xshift=-\xOff2,yshift=-\yOff2]OP2.north east) %
		{\color{red}(b2)};
		\node (OP23cap) at
		([xshift=-\xOff2,yshift=-\yOff2]OP23.north east) %
		{\color{red}(b3)};
		\end{tikzpicture}%
}

\begin{document}
\title{ Higher Order Topological Phases: A General Principle of Construction}

\author{Dumitru C\u{a}lug\u{a}ru}
\affiliation{Max-Planck-Institut f\"{u}r Physik komplexer Systeme, N\"{o}thnitzer Stra. 38, 01187 Dresden, Germany}
\affiliation{Cavendish Laboratory, University of Cambridge, J. J. Thomson Avenue, Cambridge, CB3 0HE, United Kingdom }

\author{Vladimir Juri\v ci\' c}
\affiliation{Nordita, KTH Royal Institute of Technology and Stockholm University, Roslagstullsbacken 23,  10691 Stockholm,  Sweden}

\author{Bitan Roy}
\affiliation{Max-Planck-Institut f\"{u}r Physik komplexer Systeme, N\"{o}thnitzer Stra. 38, 01187 Dresden, Germany}

\date{\today}

\begin{abstract}
We propose a general principle for constructing higher-order topological (HOT) phases. We argue that if a $D$-dimensional first-order or regular topological phase involves $m$ Hermitian matrices that anti-commute with additional $p-1$ mutually anti-commuting matrices, it is conceivable to realize an $n$th-order HOT phase, where $n=1, \cdots, p$, with appropriate combinations of discrete symmetry-breaking Wilsonian masses. An $n$th-order HOT phase accommodates zero modes on a surface with codimension $n$. We exemplify these scenarios for prototypical three-dimensional gapless systems, such as a nodal-loop semimetal possessing SU(2) spin-rotational symmetry, and Dirac semimetals, transforming under (pseudo-)spin-$\frac{1}{2}$ or 1 representations. The former system permits an unprecedented realization of a fourth-order phase, without any surface zero modes. Our construction can be generalized to HOT insulators and superconductors in any dimension and symmetry class.
\end{abstract}

\maketitle

Topological phases of matter are characterized by the topological invariant of the corresponding bulk band structure, manifesting through metallic surface states. The bulk-boundary correspondence is operative for both gapped (insulators and superconductors) as well as gapless systems. Topological invariants protecting these boundary modes may arise solely from anti-unitary (such as time-reversal, particle-hole) symmetries, as in quantum spin Hall insulators, or their combinations with unitary (such as crystalline) symmetries, as in topological crystalline insulators, or by no symmetry at all, as in quantum Hall or Chern insulators~\cite{HAS10, QI11, CHI16, ARM18, KAN05, BER06, FU06, FU07a, ZHA09, LIU10, SLA13}.

\begin{figure}[t!]
	\TKIpicture{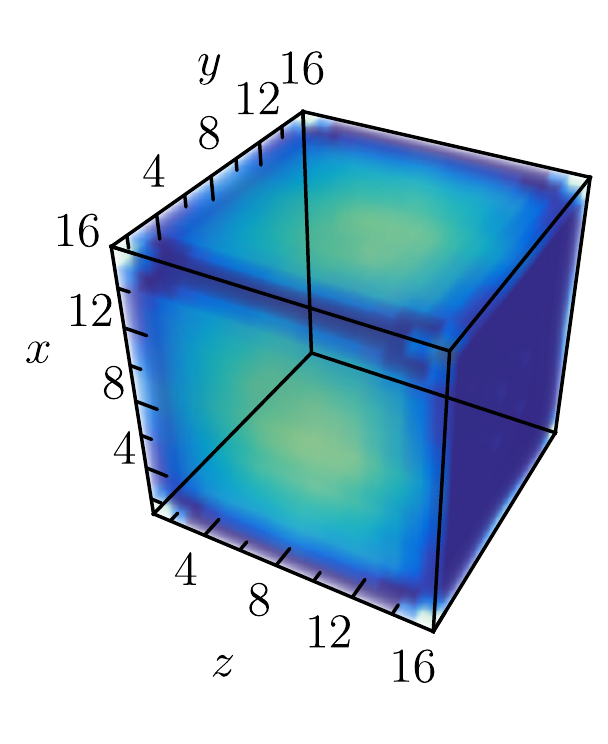}%
	{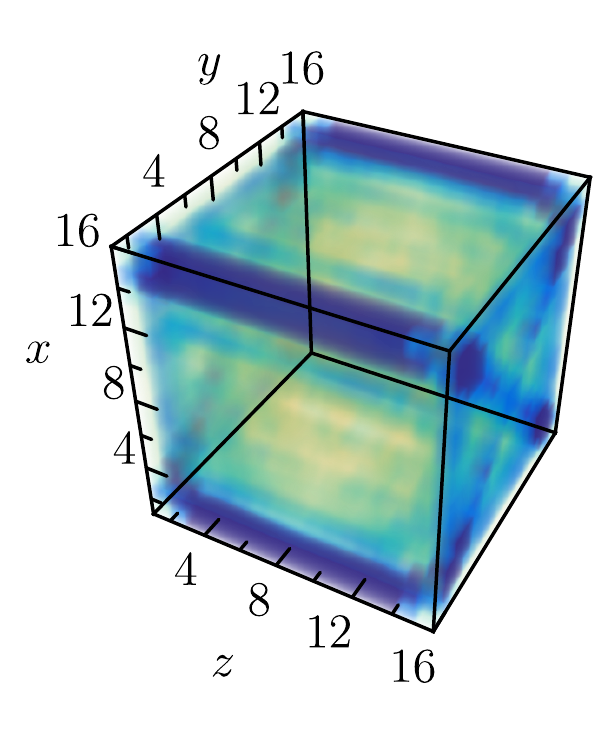}%
	{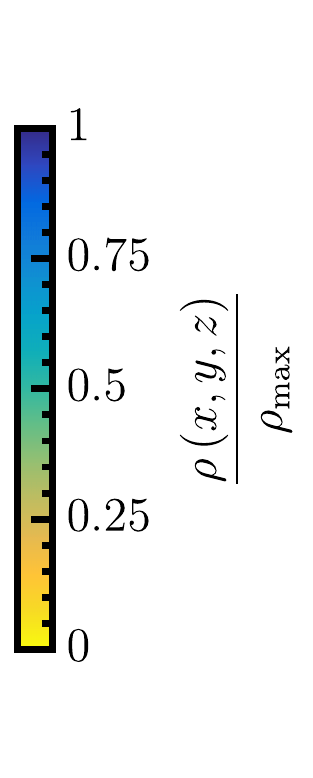}
	\caption{ Boundary modes of (a) first-order (occupying six surfaces) and (b) second-order (occupying four hinges) order topological Kondo insulators, obtained from an appropriate tight-binding model for its prototypical representative SmB$_6$~\cite{Supplementary_materials}. 
	}~\label{Fig:HOSM_HOTKI}
\end{figure}

\begin{figure}[t!]
\diracpicture{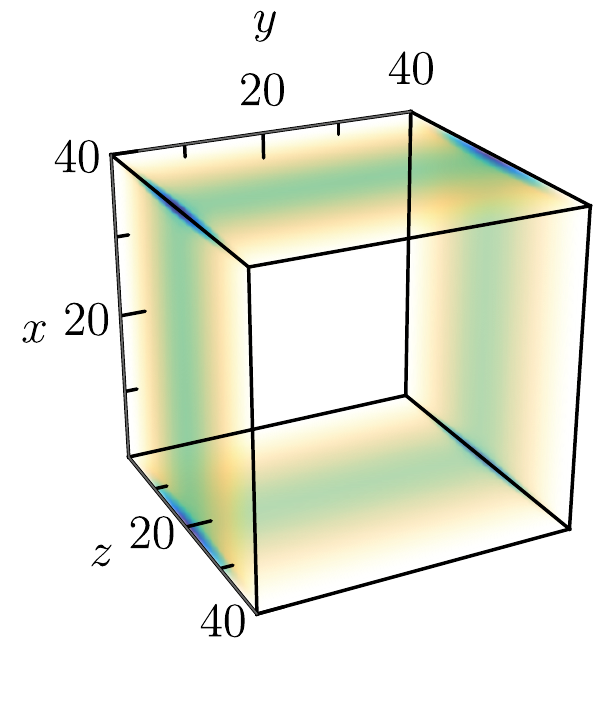}%
{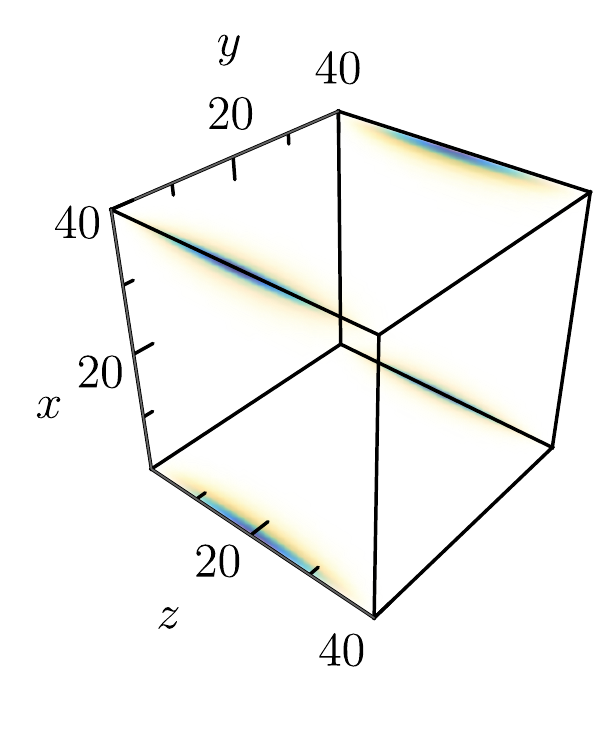}%
{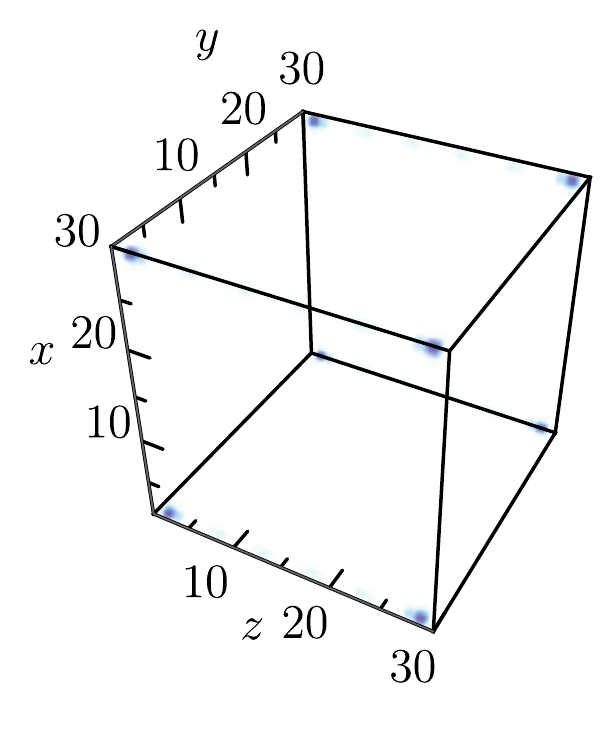}%
{colorbarParula3D_inverted}
\caption{ (a) Two-dimensional Fermi arc surface states of a first-order, (b) four one-dimensional hinge states of a second-order and (c) eight zero-dimensional corner states of a third-order spin-$\frac{1}{2}$ topological Dirac semimetals. We compute the square of the amplitude of the lowest energy [${\mathcal O}\left( 10^{-3}t \right)$ roughly] state $\rho(x,y,z)$ (normalized within the interval $[0,1]$) on a cubic lattice with open boundaries in all three directions. For numerical diagonalization we set $t=1$, $\Delta_1=\Delta_2=0.3$~\cite{Supplementary_materials}.
}~\label{Fig:HOSM_DiracHalf}
\end{figure}

\begin{figure}[t!]
\diracpicture{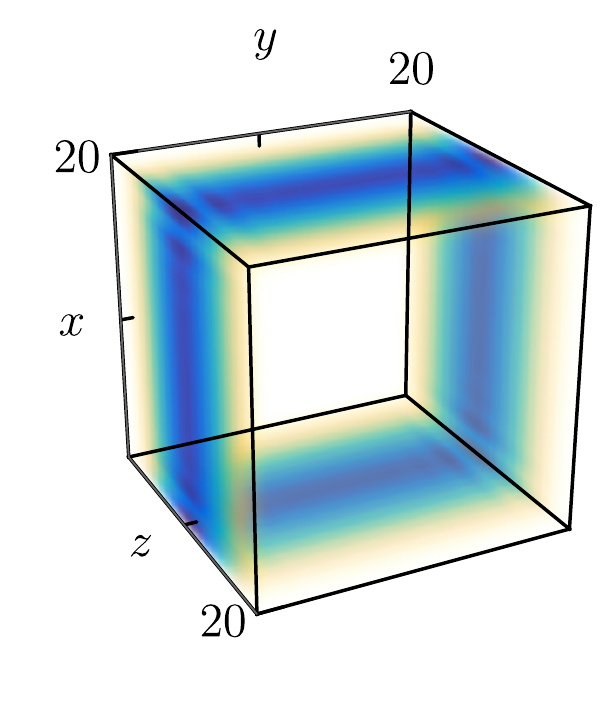}%
{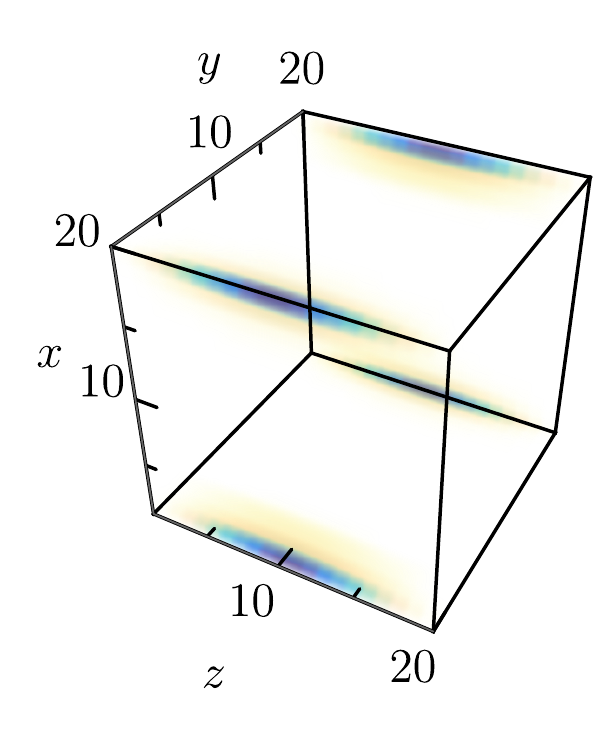}%
{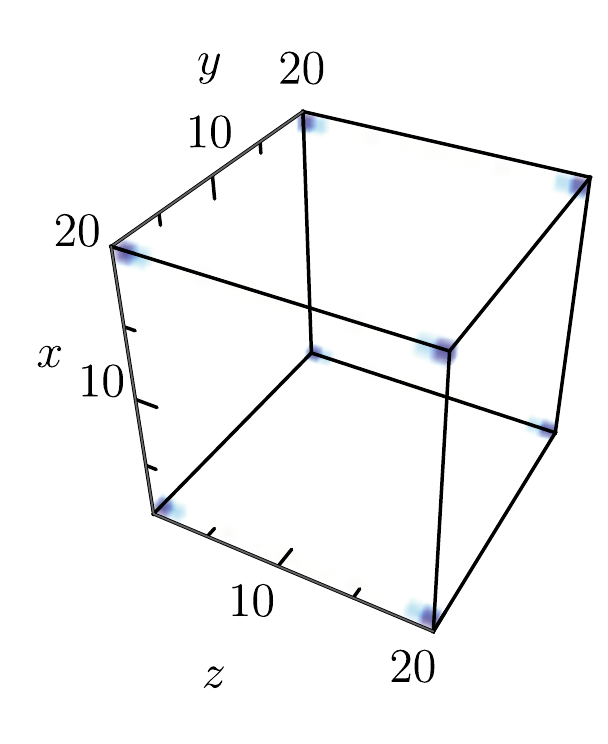}%
{colorbarParula3D_inverted}
\caption{Same as Fig.~\ref{Fig:HOSM_DiracHalf}, but for spin-1 topological Dirac semimetals. For numerical diagonalization we set $\Delta_1=0.2$ for (b), and $\Delta_1=0.1$, $\Delta_2=0.2$ for (c). In (a) and (b) [(c)] we show the low energy states maximally localized on the surface [at the corners]~\cite{Supplementary_materials}.
}~\label{Fig:HOSM_DiracOne}
\end{figure}

\begin{figure*}[t!]
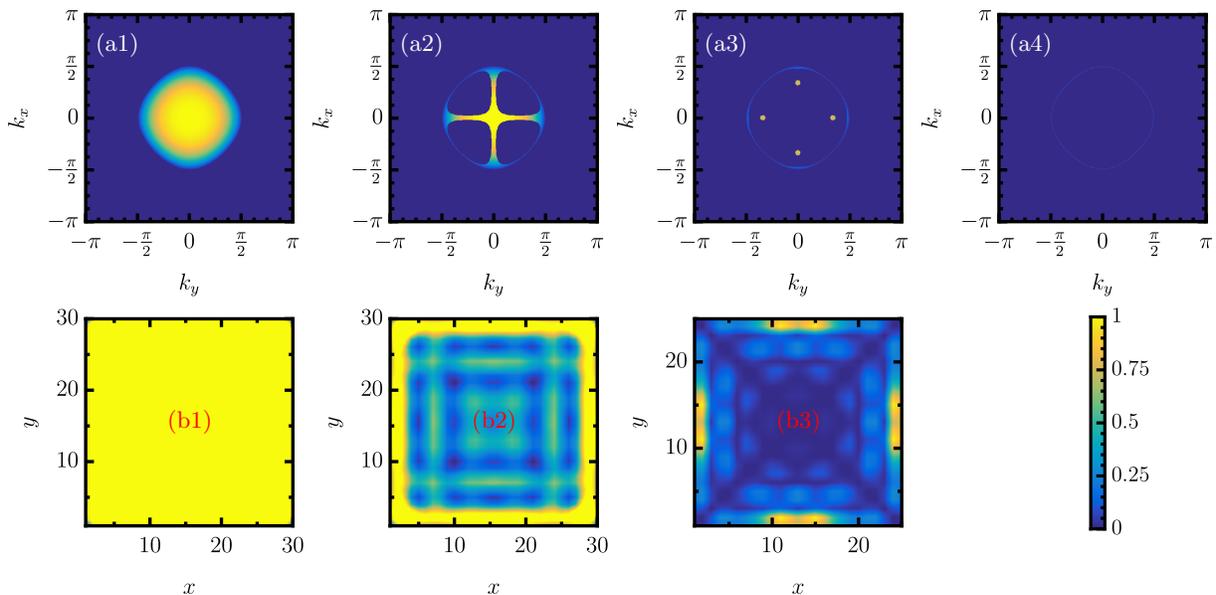

\LNSMpicture{surf_state_two_band_S_10_NOOP_mainPaper_2D}%
{surf_state_two_band_S_10_OP_2_mainPaper_2D}%
{surf_state_two_band_S_10_OP_23_mainPaper_2D}%
{surf_state_two_band_S_10_OP_123_mainPaper_2D}%
{LDOS_30_10_NOOP_2D}%
{LDOS_30_10_OP2_2D}%
{surf_state_two_band_openXYZ_S_10_OP_23_noScale_2D}%
{colorbarParula2D_mainPaper}
\caption{ Topologically protected surface modes for higher-order nodal-loop semimetals (NLSMs) in the reciprocal (top panel) and real (lower panel) space. Top panel: Square of the amplitude of near zero-energy states localized on the surface. Lower panel: Corresponding LDoS in real space [(b1) and (b2)] or amplitude of one of the near zero-energy states [for (b3)], with $E \sim  10^{-2} t$ (roughly). Here (a$n$), (b$n$) show the surface states for $n$th-order NLSM, for $n=1,2,3$, and (a4) depicts absence of any localized surface states for a $4^{\rm th}$-order NLSM. The \emph{faint} peripheral states in panels (a2), (a3) and (a4) are the shadow of the delocalized states, connecting the top and bottom surfaces through the topologically protected bulk nodal-loop. For numerical diagonalization we set $t=1$, $b=1$, $\Delta_1=\Delta_2=\Delta_3=0.3$ and $b^\prime=1.5$. The linear dimension in the $z$ direction, along which we always implement an open boundary, is $L_z=400$ (top panel), $10$ [for (b1) and (b2)] and $25$ [for (b3)]~\cite{Supplementary_materials}.
}~\label{Fig:HOSM_LNSM}
\end{figure*}

Recently, the notion of topological states of matter has been extended to include their higher-order cousins~\cite{BEN17a,PAR17}. Prototypical examples include higher-order topological (HOT) insulators~\cite{SON17,BEN17,BEN17a,LAN17,XU17,PET18,XUE18,FRA18,MAT18,SCH18,EZA18,KHA18,VAN18,SCH18a,WAN18,HSU18,EZA18a,QUE18,TRI18,Serra-Garcia2018,Thomale2018,NOH18}, superconductors~\cite{Volovik2010,KHA18,LAN17,WAN18a,TRI18,Yan2018} and semimetals~\cite{EZA18}. A $D$-dimensional $n$th-order bulk topological phase hosts gapless states on a surface with \emph{co-dimension} $d_c=n$, and bismuth (Bi) has emerged as a prominent candidate for a HOT insulator~\cite{SCH18a}. In this classification scheme, conventional topological phases are \emph{first-order} in nature. Recall that a three-dimensional ``first-order" strong $Z_2$ topological insulator supports two-dimensional gapless surface states ($d_c=1$), whereas its second-order and third-order realizations respectively accommodate one dimensional hinge modes with $d_c=2$ (see Fig.~\ref{Fig:HOSM_HOTKI}) and pointlike corner states with $d_c=3$. The present work promotes a general principle for constructing HOT phases, based on the dimensionality and symmetries of the system. For concreteness, we focus on prototypical gapless phases in three dimensions, such as Dirac and nodal-loop semimetals, and systematically construct their higher-order realizations. The central outcomes are summarized in Figs.~\ref{Fig:HOSM_DiracHalf}, ~\ref{Fig:HOSM_DiracOne}, ~\ref{Fig:HOSM_LNSM}.

To set the stage, we begin by recalling that the \emph{unstable} fixed point separating two topologically distinct gapped phases (belonging to any Altland-Zirnbauer class in any dimension~\cite{ALT97,KIT09}) is described by \emph{massless} Dirac fermions~\cite{Schnyder2010,SCH08}. The fundamentally important element in this construction is the topological Wilson mass, which introduces a nontrivial band gap for Dirac quasiparticles. Now, depending on the dimensionality and symmetry of the system, one can find several candidates for the topological mass, represented by mutually anti-commuting Dirac matrices. For example, a four-component Dirac Hamiltonian in $D=3$ (suitable to describe both topological insulators and superconductors) permits one time-reversal-symmetry ($\mathcal T$) preserving ordinary Dirac mass and a $\mathcal T$-odd \emph{axion} mass. Often the $\mathcal T$ invariant Dirac mass constitutes a first-order topological insulator or superconductor, accommodating two-dimensional gapless surface states. The presence of the additional $\mathcal T$-odd Dirac mass then either (a) completely gaps the surface states, yielding an axionic insulator~\cite{ESS09,LI10,ROY16} or superconductor~\cite{GOS14}, when it is uniform (following the trivial representation) or (b) gives birth to one-dimensional \emph{hinge} modes, when it also lacks certain discrete symmetry (such as $C_4$, hence transforming under a non-trivial representation of lattice point group). The system then describes a second-order topological insulator or superconductor. Hence, the arena of the HOT phases (including insulators, superconductors and semimetals) can be ventured in terms of available ``masses" for gapless Dirac fermions. Below we formulate a general principle of constructing a hierarchy of HOT phases.

Let us assume that the minimal model describing a first-order topological phase involves $m$ (often, but not necessarily mutually anti-commuting) Hermitian matrices. If there exist maximal $p-1 (\geq 0)$ mutually anti-commuting matrices that also anti-commute (hence represent mass orders) with the $m$ matrices, then it is conceivable to realize up to an $n$th-order HOT phase, with appropriate combinations of $n$ discrete symmetry breaking Wilsonian masses, where $n=1, \cdots, p$. For the above example of a topological insulator or superconductor in $D=3$ (described by four-component Dirac fermions) $m=4$ and $p=2$, and at most a second-order HOT phase can be realized in the presence of discrete symmetry breaking axion mass. Within this framework we construct a second order topological Kondo insulator (supporting four hinge modes) in the presence of a time-reversal symmetry breaking quadrupolar Kondo singlet mass, see Fig.~\ref{Fig:HOSM_HOTKI}(b)~\cite{Supplementary_materials}. For rest of this Rapid Communication we focus on topological semimetals and anchor this prediction for the following systems.

 (1) Topological Dirac semimetals (TDSMs), transforming under either (pseudo-)spin-$\frac{1}{2}$ or 1 representation, for which $m=3$ and $p=3$. We construct first-, second- and third-order TDSMs, respectively supporting two-dimensional Fermi arcs, one-dimensional hinge and zero dimensional corner states, see Figs.~\ref{Fig:HOSM_DiracHalf} and ~\ref{Fig:HOSM_DiracOne}. These conclusions remain operative for TDSMs, following any half-integer or integer (pseudo-)spin representation~\cite{Supplementary_materials}.

 (2) Nodal-loop semimetal (NLSM) with SU(2) spin-rotational symmetry, for which $m=2$ and $p=4$. Consequently, we can find up to fourth-order NLSM, see Fig~\ref{Fig:HOSM_LNSM}. A fourth-order NLSM does not support any surface zero modes, since $d_c>D$~\cite{Supplementary_materials}.

In the above examples of higher order topological semimetals (HOSMs), the discrete symmetry-breaking Wilson masses, the key ingredient for the realization of any HOT phase, leave the topological band touching points unchanged~\cite{Supplementary_materials}.

We begin the journey through the territory of HOSMs with TDSMs, following spin-$j$ (half-integer or integer) representation. The Hamiltonian operator is given by
\begin{eqnarray}
\hat{h}^{\rm Dirac}_j= t \left[ \tau_3 \otimes S^{j}_1 N_1({\bf k}) + \tau_0 \otimes \sum^3_{m=2} S^{j}_m N_m ({\bf k}) \right],
\end{eqnarray}
where $S^{j}_m$ are the standard $(2j+1)$-dimensional spin-$j$ matrices, $t$ is the hopping amplitude and
\begin{eqnarray}~\label{Eq:FormFactor_Dirac}
N_1({\bf k})&=& \sx, \:\: N_2({\bf k})= \sy,  \nonumber \\
N^1_3({\bf k})&=& \cz + \left[ \cx +\cy-2 \right].
\end{eqnarray}
The Pauli matrices $\{\tau_\mu \}$ operate on the (pseudo-)spin indices. For an arbitrary half-integer spin Dirac system, the energy spectra display \emph{multi-fringence} with $(2 j+1)$ effective Fermi velocities, and Kramers or (pseudo-)spin degenerate, linearly dispersing valence and conduction bands touch each other at ${\bf k}=(0,0,\pm \frac{\pi}{2 a})$, the Dirac points. Integer-spin Dirac systems additionally accommodate a pair of completely \emph{flat}, but topologically trivial, bands precisely at zero energy. For concreteness, we here focus on TDSMs following $j=\frac{1}{2}$ and $1$ representations.

TDSMs can be constructed by stacking two-dimensional quantum spin Hall insulators in the momentum space along, for example, the $k_z$ direction. Each two dimensional layer accommodates one-dimensional gapless edge modes in the $xy$-plane. The resulting TDSMs then accommodate \emph{delocalized} zero-energy surface states, but only on the $xz$ and $yz$ planes, see Figs.~\ref{Fig:HOSM_DiracHalf}(a) (for spin-$\frac{1}{2}$) and ~\ref{Fig:HOSM_DiracOne}(a) (for spin-1), yielding topologically protected Fermi arcs (localized in the reciprocal space)~\cite{SLA17, NAN18}.

Higher order TDSMs can be realized by introducing the following two discrete symmetry-breaking masses
\begin{equation}
\hat{h}^{{\rm Dirac},j}_{\rm XY-WDW} = \left[ \sum^{2}_{j=1} \; \Delta_j \: \tau_j \: N^{\prime}_j ({\bf k}) \right] \otimes M_{j},
\end{equation}
where $M_j$ is a $(2j+1)$-dimensional \emph{antidiagonal} identity matrix. In particular, we choose
\begin{eqnarray}
N^\prime_1({\bf k}) &=&\cx-\cy \sim k^2_x-k^2_y \;: \:\: {\rm B_{1g}},  \\
N^\prime_2({\bf k}) &=& \sx \sy \sz \sim k_x k_y k_z \;: \: {\rm B_{1u}}, \nonumber
\end{eqnarray}
respectively transforming under the ${\rm B_{1g}}$ and ${\rm B_{1u}}$ representations of the tetragonal point group ($D_{4h}$). These two orders mix the in-plane components of (pseudo-)spin, and we name them the XY Wilsonian density-wave (XY-WDW). Note that $N^\prime_1({\bf k})$ and $N^\prime_2({\bf k})$ respectively represent quadrupolar and octupolar orders, but both of them vanish at ${\bf k}=\left( 0,0,\pm \frac{\pi}{2 a}\right)$. Hence, they do not affect the band-touching points, but for integer-spin systems trivial flat bands acquire dispersion away from the Dirac points~\cite{Supplementary_materials}. We first introduce the ${\rm B_{1g}}$ XY-WDW.

Notice that the ${\rm B_{1g}}$ XY-WDW order acts as a \emph{mass} for the one-dimensional edge states associated with each copy of the two-dimensional spin Hall insulator, but changes sign under the $C_4$ rotation. The system can then be described as a collection of one dimensional Dirac fermions (for each slice along the $z$ direction) in the presence of a \emph{domain wall} mass. The \emph{Jackiw-Rebbi} index theorem~\cite{JAC76} then guarantees the existence of precisely one zero-energy mode at \emph{four} hinges between the $xz$ and $yz$ planes. A collection of such zero-energy modes constitutes four \emph{hinge states}, as shown in Figs.~\ref{Fig:HOSM_DiracHalf}(b) and ~\ref{Fig:HOSM_DiracOne}(b), and a second-order TDSM is realized.

The ${\rm B_{1u}}$ XY-WDW acts as a \emph{mass} for the one-dimensional hinge states of the second-order TDSM, but assumes the profile of a \emph{domain-wall} along the $z$ direction. Therefore, a subsequent addition of the ${\rm B_{1u}}$ XY-WDW causes further dimensional reduction of the the hinge modes, giving rise to eight \emph{corner} states [see Figs.~\ref{Fig:HOSM_DiracHalf}(c) and \ref{Fig:HOSM_DiracOne}(c)] and a third-order TDSM is realized.

We now focus on a NLSM possessing an exact SU(2) spin rotational symmetry. The Hamiltonian operator is given by
\begin{eqnarray}~\label{Eq:NLSM_unperturbed}
\hat{h}^{\rm NL}_0= \sigma_0 \otimes \left[ t\; \sum^{2}_{i=1} N_i(\mathbf{k}) \; \tau_j \right].
\end{eqnarray}
Two sets of Pauli matrices $\left\{ \sigma_\mu \right\}$ and $\left\{ \tau_\mu \right\}$ respectively operate on the spin and sublattice indices. The spinor basis is $\Psi^\top_{\bf k}=\left( \Psi_{{\bf k}, \uparrow}, \Psi_{{\bf k}, \downarrow} \right)$, with $\Psi^\top_{{\bf k}, \sigma}=( c^{\rm A}_{{\bf k},\sigma}, c^{\rm B}_{{\bf k},\sigma} )$ for two projections of electron spin $\sigma=\uparrow, \downarrow$. Here $c^{\rm X}_{{\bf k},\sigma}$ is the fermionic annihilation operator on sublattice $\rm X=A,B$, with momentum ${\bf k}$ and spin projection $\sigma$. We choose~\cite{ROY17}
\allowdisplaybreaks[4]
\begin{eqnarray}
	  N^1_1(\mathbf{k}) &=& \cx+\cy-b, \nonumber \\
		N^2_1(\mathbf{k}) &=& \cz-1, \:\: N_2(\mathbf{k})= \sz,
\end{eqnarray}
with $N_1({\bf k})=N^1_1({\bf k})+N^2_1({\bf k})$ and $b<2$. Here $N^2_1({\bf k})$ plays the role of a regular Wilson mass (preserving all symmetries) that removes the nodal loops at $k_z=\pm \frac{\pi}{a}$, leaving the one at $k_z=0$ untouched. The inversion symmetry ${\mathcal P}$, under which ${\bf k} \to - {\bf k}$ and $\Psi_{\bf k} \to \tau_1 \Psi_{-\bf k}$ ensures the stability of the nodal-loop in the $k_z=0$ plane. The anti-commutation relation $\left\{\tau_3, \hat{h}^{\rm NL}_0 \right\}=0$ guarantees the spectral symmetry and spin-degenerate \emph{drum-head} surface states are pinned at precisely zero energy on two opposite surfaces along the $(001)$ direction, see Fig.~\ref{Fig:HOSM_LNSM}(a1). The corresponding local density of states (LDoS) in the $xy$-plane (top surface) is completely uniform, see Fig.~\ref{Fig:HOSM_LNSM}(b1).

Higher order NLSMs can be realized by invoking the Wilsonian spin-density-wave (WSDW) mass orders, captured by the effective single-particle Hamiltonian
\begin{equation}~\label{Eq:WSDW}
\hat{h}^{\rm NL}_{\rm WSDW}= N^2_1({\bf k}) \left[ \sum^{3}_{j=1} \Delta_j \: \sigma_j \: \otimes N^\prime_j(\mathbf{k})  \right].
\end{equation}
Three components of WSDW are chosen to be
\begin{align}~\label{Eq:WSDW_NLSM}
		 N'_1(\mathbf{k})  &= \Sx \Sy \tau_3            \sim  k_x k_y \tau_3 : {\rm B_{1u}} \nonumber \\
		 N'_2(\mathbf{k})  &= [\Cx+\Cy-b^\prime] \tau_3 \sim (k^2_\perp -b^\prime) \tau_3: {\rm A_{2u}}, \nonumber \\
		N'_3(\mathbf{k})  &= [\Cx-\Cy] \tau_3          \sim (k^2_x -k^2_y)\tau_3 : {\rm B_{2u}},
\end{align}
where $k^2_\perp=k^2_x+k^2_y$. We expand ${\bf N}'_j({\bf k})$ around the center of the Brillouin zone, allowing us to classify the WSDW orders according to their transformation under the $D_{4h}$ group. Notice that all WSDW orders vanish at $k_z=0$, due to the appearance of $N^2_1({\bf k})$ in $\hat{h}^{\rm NL}_{\rm WSDW}$ [see Eq.~(\ref{Eq:WSDW})]. Thus the bulk nodal-loop residing in the $k_z=0$ plane remains protected in the presence of any such order~\cite{Supplementary_materials}.

The ${\rm B_{1u}}$ WSDW order reduces the two-dimensional drumhead surface state (with $d_c=1$) into a \emph{pair} of one-dimensional \emph{arc states}, given by $k_x=0$ and $k_y=0$ in the $\left( k_x,k_y\right)$ plane, see Fig.~\ref{Fig:HOSM_LNSM}(a2). Across these two specific directions the ${\rm B_{1u}}$ order changes its sign, and takes the profile of a domain wall mass. The corresponding LDoS in the $xy$ plane reveals the underlying one-dimensional hinge modes (with $d_c=2$), see Fig.~\ref{Fig:HOSM_LNSM}(b2), realizing a second-order NLSM.

The subsequent addition of the ${\rm A_{2u}}$ WSDW order introduces a mass order for one-dimensional arc states of the second-order NLSM, with the profile of a domain wall. Specifically, this order changes sign when $\cos(k_j a)=b^\prime$ for $j=x,y$. The surface states then undergo additional dimensional reduction and now support only four isolated zero-dimensional states in the $\left(k_x,k_y\right)$ plane, see Fig.~\ref{Fig:HOSM_LNSM}(a3), when $b^\prime>b$. The spatial distribution of one such state in the $xy$-plane is shown in Fig.~\ref{Fig:HOSM_LNSM}(b3), which takes the shape of zero-dimensional corner states (with $d_c=3$), but now localized at the center of four edges. The resulting phase represents a third-order NLSM.

Finally, when we include ${\rm B_{2u}}$ WSDW order the surface modes become completely gapped, see Fig.~\ref{Fig:HOSM_LNSM}(a4), and a fourth-order NLSM is realized. For all realizations of higher-order NLSMs the perimeter of the original drumhead surface state, accommodating \emph{delocalized} states connecting the top and bottom surfaces through the bulk nodal-loop, remains unaffected. In turn this observation confirms that the WSDW orders (a) do not affect the bulk nodal-loop, and (b) only cause dimensional reduction of the surface states, the quintessential features of any HOT phase of matter. The existence of the topologically protected bulk nodal-loop for all higher-order NLSMs is further substantiated from the scaling of the density of states, namely $g(E) \sim |E|$ in the bulk~\cite{Supplementary_materials}.

We note that it is also conceivable to realize a second-order NLSM in the presence of a Wilsonian charge-density-wave, obtained by taking $\sigma_j N^\prime_j ({\bf k}) \to \sigma_0 N^\prime_1 ({\bf k})$ in Eq.~(\ref{Eq:WSDW}). The resulting arc states in the momentum space and the hinge modes in the real space are similar to the ones shown in Figs.~\ref{Fig:HOSM_LNSM}(a2) and (b2), respectively, but, we cannot proceed through the hierarchy of HOSMs any further with the charge-density-wave order.


To summarize, we here present a general principle for constructing HOT phases, including gapped~\cite{Supplementary_materials} and gapless representatives. In particular, we illustrate the higher-order generalization of a topological Kondo insulator (see Fig.~\ref{Fig:HOSM_HOTKI}), TDSMs (see Figs.~\ref{Fig:HOSM_DiracHalf} and \ref{Fig:HOSM_DiracOne}), which can be realized in Cd$_3$As$_2$~\cite{BOR14}, Na$_3$Bi~\cite{LIU14}, and a NLSM (see Fig.~\ref{Fig:HOSM_LNSM}), relevant for Ca$_3$P$_2$~\cite{XIE15}, when subjected to lattice deformations thereby reducing the symmetry. This construction can also be applied to gapless superconductors, where the zero-energy modes residing on a surface of co-dimension $d_c=n$ are constituted by \emph{Majorana} fermions. In particular, localized Majorana zero modes in the form of the \emph{corner states} can be useful for topological quantum computations~\cite{NAY08}.

B. R. and V. J. are thankful to Titus Neupert for useful discussions. B. R. is thankful to Nordita for hospitality during the workshop ``Topological Matter Beyond the Ten-Fold Way" where this work was initiated.

\bibliographystyle{apsrev4-1}
\bibliography{Higher_Order_Topology,notes}

\end{document}